\documentclass[10pt,conference]{IEEEtran}
\makeatletter
\def\@author{}
\makeatother
\usepackage[labelfont=bf,textfont={bf}]{caption}
\usepackage{textcomp}
\usepackage{dblfloatfix}  
\usepackage{float}
\usepackage{balance}
\usepackage{cite} 
\usepackage{algorithmic}
\usepackage{graphicx} 
\usepackage{subcaption}
\usepackage{alltt}
\usepackage[hidelinks]{hyperref} 
\usepackage{url} 
\usepackage{rotating}
\usepackage{adjustbox}
\usepackage{booktabs}
\usepackage{enumitem}
\usepackage{algorithm}
\usepackage{algorithmic}
\usepackage{lineno}
\usepackage{mdframed} 
\usepackage{fontawesome5}  
\usepackage{amssymb}       
 \usepackage{amsmath} 
\usepackage{graphicx} 
\usepackage{array}    
\usepackage[table]{xcolor} 
\usepackage{multirow}
\usepackage[most]{tcolorbox} 
\usepackage{wrapfig}
\setitemize{noitemsep,topsep=0pt,parsep=0pt,partopsep=0pt}
\newcommand{\bi}{\begin{itemize}}
\newcommand{\ei}{\end{itemize}}

\newlength{\unitlengthforbar}
\definecolor{grayline}{gray}{0.8}
\definecolor{lightyellow}{RGB}{255,255,204}
\definecolor{lightpurple}{RGB}{229,204,255}
\definecolor{lightred}{RGB}{255,204,204}
\definecolor{lightgrey}{RGB}{224,224,224}
\definecolor{lightblack}{RGB}{82,82,82} 
\newenvironment{conclusion}[1]
{
  \begin{mdframed}[
    leftline=true,               
    topline=false,               
    bottomline=false,            
    rightline=false,             
    linecolor=black,             
    linewidth=4pt,               
    backgroundcolor=gray!20,     
    innertopmargin=6pt,          
    innerbottommargin=6pt,       
    skipabove=10pt,              
    skipbelow=10pt,              
    innerleftmargin=6pt,        
    rightline=false,             
  ]
  \textbf{#1} 
  \ignorespaces
}
{\end{mdframed}
}  

\newcommand{\IT}{\textsf{{\sffamily BINGO}}}
 \newcounter{myquotecounter}

\begin{document}
\title{ BINGO! Simple Optimizers Win Big if Problems Collapse to a Few Buckets  }
\author{
  Kishan Kumar Ganguly, and Tim Menzies \\
  Department of Computer Science, North Carolina State University, Raleigh, USA\\
  kgangul@ncsu.edu, timm@ieee.org
}
\IEEEtitleabstractindextext{
\begin{abstract} 

Traditional multi-objective optimization in software engineering (SE) can be slow and complex. This paper introduces the {\sffamily BINGO} effect: a novel phenomenon where SE data surprisingly collapses into a tiny fraction of possible solution "buckets" (e.g., only 100 used from 4,096 expected).

We show the {\sffamily BINGO} effect's prevalence across 39 optimization in SE problems. Exploiting this, we optimize 10,000
  times faster than state-of-the-art methods, with comparable effectiveness. Our new algorithms (LITE and LINE), demonstrate that simple stochastic selection can match complex optimizers like DEHB. This work explains why simple methods succeed in SE—real data occupies a small corner of possibilities—and guides when to apply them, challenging the need for CPU-heavy optimization. 
  
  Our data and code are public at  GitHub (anon-artifacts/bingo).
  
\end{abstract}

}

\maketitle
\IEEEpeerreviewmaketitle
\IEEEdisplaynontitleabstractindextext

\begin{table*}[!t]
\caption{39 SE optimization problems. {\em x/y}  denotes
number of inputs and output  goals. For more details,  see Table~\ref{details}. }
\label{combinedtable}
\scriptsize
\centering
\begin{tabular}{>{\raggedright\arraybackslash}m{0.5in}>{\raggedright\arraybackslash}m{1in}>{\raggedright\arraybackslash}m{1.2in} >{\raggedright\arraybackslash}m{2.6in} >{\raggedright\arraybackslash}m{0.5in}}
\textbf{\# Data sets} & \textbf{Dataset Type / Groups} & \textbf{File Name / Features} & \textbf{Primary Objective} & \textbf{$x/y$} \\
 \hline

25 &  Specific Software Configurations & SS-A to SS-X, billing\_10k & Optimize software system settings e.g. Runtimes, query times, and usage data from software configured in various manners & 3-88/2-3
\arrayrulecolor{grayline} \\\hline
1 &cloud & Apache AllMeasurements & Apache server performance optimization & 9/1 \\
1 &cloud    & SQL AllMeasurements & SQL database tuning & 38/1 \\
1 & cloud   & X264 AllMeasurements & Video encoding optimization & 16/1 \\
7 &   cloud   & (rs|sol|wc)* & misc configuration tasks  & 6/1 \\\hline
2 & Health Data Sets & healthCloseIssues12mths & Predict project health and developer activity & 5/3 \\\hline
1 & Debug Datasets & auto93 & Misc  & 4/3 \\\hline
1 & Miscellaneous & HSMGP num & Hazardous Software Management Program data & 14/1  \\\hline
39 & Total
\end{tabular}
\vspace{-5mm}
\end{table*}
\section{Introduction}

Many SE problems use multi-objective optimization (e.g., maximizing features for minimal cost \cite{Sayyad:2013} or throughput for minimal energy \cite{Apel2020}), as shown by Sarro \cite{sarro2023search}. Traditional methods, like simulated annealing \cite{Menzies07a} or genetic algorithms \cite{linares2015optimizing}, are slow due to evaluating thousands to millions of solutions; e.g. Zuluaga et al. \cite{zuluaga2013active} noted that ``synthesis of only one (solution) can take hours or even days''. Such runtimes impede reproducibility, as statistical validity for papers requires multiple runs across many datasets. 
Routinely, 
the experiments required for our papers take days to weeks (and sometimes, months). 
While these runs can be parallelized, we often find that the cost   of     those executions  is    a hard limit on  progress.

This paper investigates the following  previously unreported phenomenon— which suggests that for optimization in SE, CPU-heavy methods may     be unnecessary:
\begin{conclusion}{The {\sffamily BINGO} effect}
When SE data is split into $n$ buckets across $d$
dimensions divided into $b$ bins, most data goes to surprisingly few
buckets; i.e. $n \ll b^d$.
\end{conclusion}
For example, in one  study with 10,000 rows, we might
expect $b^d=4096$ buckets from $d=4$ dimensions split into $b=8$ bins.
Yet in practice, we find only $100$ used buckets.

Evaluating millions of options is unnecessary when only (say)
100 distinct scenarios exist. But how common and useful is
this {\sffamily BINGO} effect? To answer that, we pose four questions.

{\bf RQ1}: {\em Is the {\sffamily BINGO} effect prevalent?}
 {\sffamily BINGO}
exists in all 39 of the SE optimization  tasks of     Tables~\ref{combinedtable},
\ref{details}.

{\bf RQ2}: {\em Is {\sffamily BINGO}-based reasoning effective?}
Compared to top optimizers (e.g. DEHB \cite{awadijcai2021p296}),
we  perform as well, or better.

{\bf RQ3}: {\em Is {\sffamily BINGO}-based reasoning fast?}
Compared to state-of-the-art methods, {\sffamily BINGO} can optimize 10,000 times faster.

{\bf RQ4}: {\em Is {\sffamily BINGO}-based reasoning simple?}
Often, simple stochastic selection suffices. When data
 is costly, we recommend  two new lightweight algorithms, LITE and LINE



What is the novelty of this analysis? Other SE researchers have reported the benefits of 
data pruning\cite{yu2019improving, tu2021frugal,
majumder2024less, kocaguneli2012active, chen2005finding,
menzies2008implications},  
 grounded perhaps in  the {\em manifold assumption}\footnote{High-dimensional
data often lies on a simpler, lower-dimensional "manifold" or surface
embedded within that higher space \cite{zhu2005semi}.} or the 
{\em Johnson-Lindenstrauss lemma}\footnote{High-dimensional points
can be projected into lower dimensions while nearly preserving all
pairwise distances \cite{johnson1984extensions}.}. But that work achieved comparatively modest reductions: Hall and Holmes
\cite{hall2003benchmarking} only removed $\approx$50\% of columns. 
Zhu~\cite{yu2019improving}  only pruned 70–80\% of rows. Our
reductions are far more extensive.

Also, those  other studies  
used limited datasets and outdated algorithms.  Agrawal et
al. \cite{agrawal2019dodge} ran against a 1980s genetic
algorithm. Lustosa \cite{lustosa2023optimizing} only studied a handful of problems, and ran against an algorithm from 2012 (Hyperopt~\cite{bergstra2012random}), that has since  been superseded by  the DEHB algorithm\cite{awadijcai2021p296} (which we use here). Hence we assert that our
use of a large corpus and state-of-the-art algorithms fills a critical
research gap.

Further,   prior work   showed  data  pruning's   efficacy   without
explaining \emph{why} it worked or \emph{when not} to use it.
\sffamily BINGO\normalfont, on the other hand explains why we can prune so much: 
 \begin{conclusion}{}
 Real data lives in a tiny corner of all possibilities.
 \end{conclusion}
 \sffamily BINGO \normalfont  also  tells us when simpler methods are not appropriate: \underline{\bf do not} use our simpler approaches if data collapses
to many thousands of bins, or more.

 Overall, our  new contributions  are as follows. This paper:
\begin{enumerate} \item Compares  stochastic and state-of-the-art   SE optimizers.
\item Confirms {\sffamily BINGO} exists in   many SE optimization tasks.
\item Shows   simple tools  achieve state-of-the-art performance.
\item Introduces LITE, a new algorithm for  $\le 24$ labels.
\item Introduces LINE, a new algorithm for   $\le 50$ labels.
\item Explains the success of simple methods like LITE and LINE: many   problems collapse to just a few   buckets.
\item Proposes a test for when to avoid simpler methods; i.e. do not try simple if data has too many buckets.
\item Shares all  the  scripts and data used in this analysis\footnote{\url{http://github.com/anon-artifacts/bingo}}.
\end{enumerate}
The rest of this paper  details {\sffamily BINGO} and its relevance to SE optimization (in the next section). Then we describe our data, algorithms, and statistics,
followed by results showing state-of-the-art tools do no
 better than simpler ones (explained via {\sffamily BINGO}). Discussion
and threats to validity follow.

 
Before beginning, we address a common question: what role do
deep learning models—specifically Large Language Models (LLMs)—
play in this kind of analysis? Prior work by Senthilkumar et al.
\cite{senthilkumar2024can} explored using LLMs to warm-start
optimization by generating strong initial guesses. They found that  even for
small subsets of our data, such approaches required days of CPU
time. Moreover, as the dimensionality of the optimization task
increased, the quality of LLM-generated solutions degraded. Other
LLM studies \cite{somvanshi2024survey} similarly report that
models with trillions of parameters struggle with tabular data such as
Table~\ref{combinedtable}. These models are costly to run and often
unsuited for optimization tasks that require rapid, scalable inference
over structured, low-signal tabular spaces. Hence, for the problems
considered here, simpler methods—not deep learning—can be more
effective.

\begin{figure}[!b] 
\begin{center} \includegraphics[width=\linewidth]{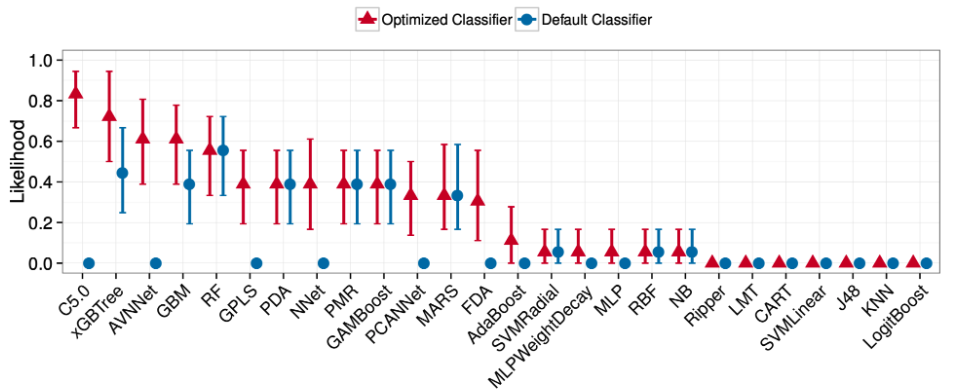}\end{center}
\caption{Defect prediction: likelihood of a learner performing
best. BLUE shows performance pre-optimization and RED shows
very large performance improvements (post-optimization). E.g.
on the left-hand-side, the C5.0 decision tree learner had the {\em worst} performance (pre-optimization) and {\em best}
performance (post-optimization). From~\cite{Tantithamthavorn16}.}
\label{tant}
\end{figure}
\section{Why Study Optimization in SE?}
This paper is about SE and  optimization-- specifically   hyperparameter optimization (HPO) and automated
configuration. We study optimization in SE since it is often poorly executed in industry.
For industrial data miners, poorly-chosen
hyperparameters  often  yield sub-optimal results~\cite{rahul16fse}.
HPO explores and improves learner configurations to
boost SE model performance~\cite{fu2016tuning,
agrawal2019dodge, yedida2021value, yedida2023find}. Figure~\ref{tant} offers one example of the dramatic improvements that can be achieved via HPO.

HPO is a special case of a more general problem.
\textbf{Software configuration} involves selecting components, parameters,
and features to tailor a system for its specific operating environment or task.
Exploring a system's configuration space $\mathcal{C}$ is an
optimization problem~\cite{Siegmund2015}. The goal is to find
$c^* \in \mathcal{C}$ that optimizes objectives (e.g.,
throughput). If $f: \mathcal{C} \rightarrow \mathbb{R}^M$ maps
a configuration to performance metrics, we seek:
\[c^* = \operatorname{argmax}_{c \in \mathcal{C}} f(c)\]
The problem here is that 
$|\mathcal{C}|$ can be astronomically large; 460 binary choices
(MySQL, 2014) yield $2^{460}$ options---vastly exceeding the
$2^{80}$ stars in the universe~\cite{doe2023personal}. 

Because the configuration problem is so large, it is often poorly managed.
Most software systems
have numerous, ill-documented, configuration choices with subtle
interactions. Default settings often prove unreliable, manual tuning can be unscalable, and configuration via human intuition is often fallible~\cite{nair2018finding}. As
systems grow, managing this complexity and its trade-offs (e.g.,
{\em more} code for {\em less} cost, or {\em faster} queries using {\em less} energy)
is difficult~\cite{Xu2015}.

\begin{table}[!t]
\caption{Details on Table~\ref{combinedtable}.}\label{details}
\begin{tabular}{|p{.95\linewidth}|}\hline
\rowcolor{blue!10}
The SS-* data sets  were collected by running software configured in different ways (selected at random) and then collecting various performance statistics (runtimes, query times, usage, etc.). The goal of these datasets is to find the software configuration that best reduces the overall software goals of each specific project.
\\
The HSMGP-*,  SQL-*, and X264-*
  datasets are similar to the SS-Models, the only difference is that instead of being related to random non-descript software configurations they are linked to specific pieces of software. Specifically Apache configurations, HSMGP configurations, SQL configurations, and configurations to the X264 encoding software.
\\\rowcolor{blue!10}
The RS-*,  SOL-*, WC-* datasets are like  SS-* data
but come from different configurations of (e.g.)
cloud-based word count tools..
\\\
 Finally, we include a simple car design data set
 (auto93) since this is useful in explaining MOOT and optimization to a non-SE audience.\\\hline
\end{tabular}
\end{table}

\begin{figure*}[!t]
\centering

\begin{minipage}{3in}
\centering
Figure~\ref{badchoices}a: software accumulates config.
\end{minipage}
\hspace{0.15in}
\begin{minipage}{2in}
\centering
Figure~\ref{badchoices}b: much config is ignored
\end{minipage}

\vspace{0.5em}

\begin{minipage}{3in}
\centering
\includegraphics[width=3in]{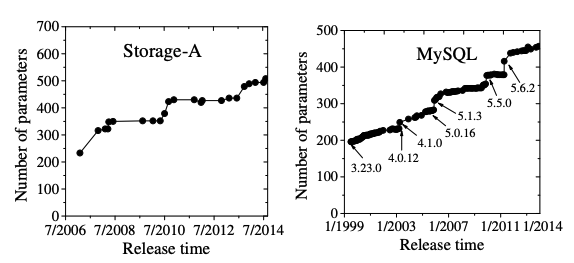}
\end{minipage}
\hspace{0.15in}
\begin{minipage}{2in}
\centering
\includegraphics[width=2in]{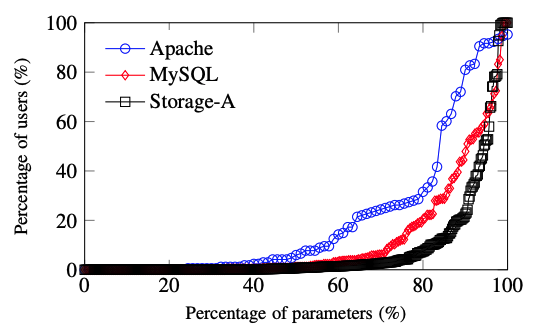}
\end{minipage}

\caption{
Number of configuration parameters over time (left).
Usage of those parameters (right). From~\cite{Xu2015}. }
\label{badchoices}
\end{figure*}

Not only is configuration a problem, but so are
configuration errors. Misconfiguration is a primary source of
software failure~\cite{Zhou2011}. Zhou et al. found that over 40\%
of MySQL, Apache, and Hadoop failures stemmed from configuration
errors, observing that such errors significantly contribute to
outages and are hard to avoid, detect, and debug. These
failures also arise from neglect; for example, MySQL's 2016
defaults used obsolete assumptions (e.g., 160MB
RAM)~\cite{VanAken2017}, standard Hadoop settings for text
mining led to worst-case execution times~\cite{Herodotou11},
and Apache Storm's worst configuration yielded 480x slower throughput
than its best~\cite{Jamshidi2016proposal}.

In Figure~\ref{badchoices}.a, Xu et al.~\cite{Xu2015} document
 a  configuration crisis  where
systems  progressively  gain  more and more configuration options, which are
mostly ignored by users.  Figure~\ref{badchoices}.b shows that for Apache/MySQL/Storage-A, 80\% of the parameters are ignored by
70-90\% of users. Van Aken et
al.~\cite{VanAken2017} confirm this, noting PostgreSQL/MySQL
configuration options increased 3x/6x over 15+ years. This
``configuration gap'' between capability and actual use yields
sub-optimal systems.

The complexity of configuration spaces, often dynamic with feature
interactions and dependencies, demands automated techniques to
reason about performance and correctness~\cite{Apel2020}. As Apel
states, without automation, "the complexity of configuration
spaces will overwhelm developers and users alike."  
Exhaustive search is impractical. Thus, intelligent sampling and machine
learning are vital tools to efficiently explore this space for
optimal configurations.

\subsection{Example Problems for Optimization in SE}
Shared datasets from research papers by Apel~\cite{Apel2020}, Chen~\cite{Li22}, and Menzies~\cite{nair2018finding}
are often used as case studies of  optimization in SE
research papers. 
 Chen and Menzies are collaborating to curate
the MOOT repository (Multi-Objective
Optimization Testing\footnote{\url{http://github.com/timm/moot/optimize}}) which offers datasets from recent SE optimization papers
for process
tuning, DB configuration, HPO, management decision making etc.

Since our focus is on configuration, we use MOOT  data related to that task (see Table~\ref{combinedtable} and~\ref{details}).
Fig.~\ref{moot}
shows the  typical structure  of those MOOT data sets. The goal in this data is to tune {\em Spout\_wait, Spliters, Counters} in order to achieve the best {\em
Throughput/Latency}. In summary:
\begin{itemize}
\item
MOOT datasets
are tables with $x$ inputs and $y$ goals.
\item The first row shows the column names. 
\item Numeric columns start with uppercase, all others are symbolic.
\item 
Goal columns (e.g. Fig.~\ref{moot}'s {\em
Throughput+, Latency-}) use +/- to denote  maximize and minimize.
\end{itemize}
 Note that our  data is much larger than the Table~\ref{moot} example.
The 39 data sets in Table~\ref{combinedtable}
have up to  86,000 rows, 88 independent variables,
and three $y$ goals.

For the purposes of illustration, the rows in Table~\ref{moot}  are sorted
from best to worst based on those goals. During experimentation, row order should initially be randomized.

For the purposes of evaluation,
all rows in  MOOT data sets contain all their $y$ values. When evaluating the outcome of
an optimizer, these values are   used to determine how well
the optimizer found the best rows. 

\begin{figure}[ht!]
\caption{An example MOOT dataset
(skipping middle rows).}\label{moot}
{\scriptsize
\begin{alltt}
  x = independent values          | y = dependent values
  --------------------------------|----------------------
  Spout_wait, Spliters, Counters, | Throughput+, Latency-
     10,        6,       17,      |    23075,    158.68
      8,        6,       17,      |    22887,    172.74
      9,        6,       17,      |    22799,    156.83
      9,        3,       17,      |    22430,    160.14
    ...,      ...,      ...,           ...,    ...
  10000,        1,       10,      |   460.81,    8761.6
  10000,        1,       18,      |   402.53,    8797.5
  10000,        1,       12,      |   365.07,    9098.9
  10000,        1,        1,      |   310.06,    9421
\end{alltt}
}
\vspace{-2mm}

\end{figure}
For the purposes of optimization experiments, researchers should hide the $y$-values from the optimizer. Each time the optimizer requests the value of a particular row, this ``costs''  one unit. For reasons described
below,  good optimizers find good goals at least cost (i.e. fewest labels).
 
\begin{table*}[!t]
 \caption{Data can be divided by adding $bins$ per dimension, or adding dimensions $d$. On the left, increasing  data  divisions  increases the number
 of   buckets with data. On the right, it does not (since the data is more clumped).}\label{eg1234}
 \begin{center}
 \includegraphics[width=4in]{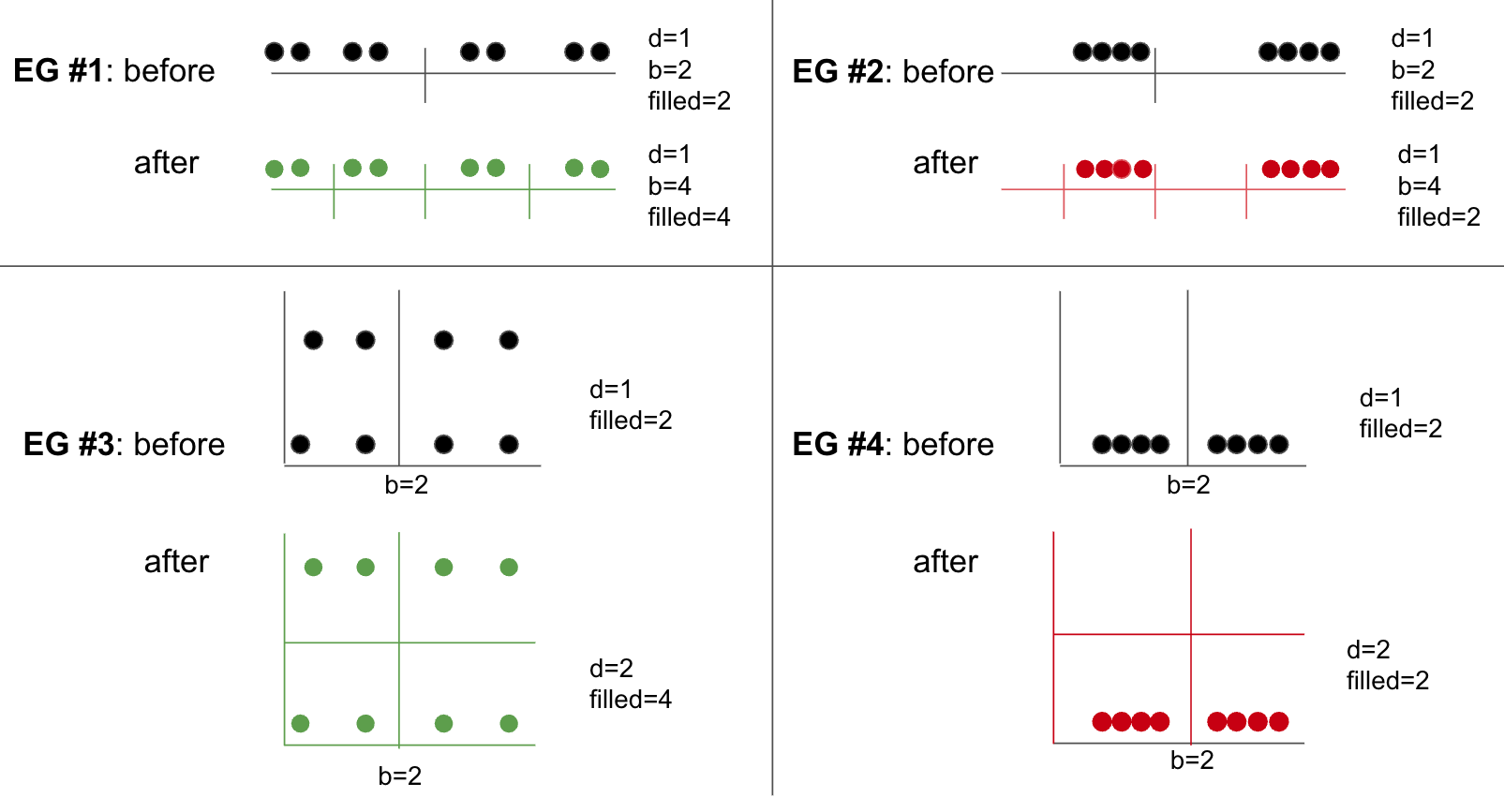}
 \end{center}
 \vspace{-4mm}
 \end{table*}
\subsection{The Labeling Problem}\label{hardlabel}
There are many  reasons to explore solutions via minimal $y$ labeling.
A common SE situation is that 
 collecting
independent $x$ variables  is often much cheaper than collecting
dependent $y$ values. For example,
\begin {itemize}
\item
  Unlabeled $x$ data (e.g., GitHub
code) is abundant but high-quality labeled $y$ data (e.g.,  team
effort to create software) is scarce and hard to obtain. 
\item
It is relatively easy to list
 many configuration possibilities within a Makefile.
 But stakeholder vetting,  or full testing of all those possibilities, can be very slow.
 \item
 With grammar-based fuzzing, it can be fast to  generate test case inputs but very slow to
run all those tests. It can be even slower for a human to check through all those results
looking for  anomalous behavior.\\
 \end{itemize}
 Since collecting $y$ can be so expensive, it is prudent to seek good rows after sampling the least $y$ labels.

 Another reason to explore minimizing $y$ labeling is that many labels are untrustworthy.
SE data can be both voluminous and dubious.
 In defect prediction data sets, 
 commits might be  labeled  ``bug-fixing'' if they contain
keywords like ``bug, fix''~\cite{catolino2017just,
hindle2008large, kamei2012large, kim2008classifying,
mockus2000identifying}, an ad hoc process cautioned by
Vasilescu et al.~\cite{vasilescu2018personnel,
vasilescu2015quality}. High rates of mislabeling occur:
over 90\% of ``false positives'' in technical debt work
~\cite{9226105}, and many errors in security bug
labeling~\cite{9371393} and static analysis false alarm
identification~\cite{10.1145/3510003.3510214}. 
Flawed data sets
need curation. Curation is  slow and expensive
{\em unless} it only needs to look at a few labels in a few rows.

For all these reasons, this paper explores SE optimization under the assumption that   very few labels can be accessed. 
Accordingly, our optimizers have a {\em budget}
of how many $y$ labels they can access. Initially,
that budget is set     small, then   is increased; e.g. 
\[
\mathit{budget}=\{6,12,18,24,50,100,200,...\} \]
  Budget   increases stop when double the budget does no better than half that number.





 \section{  Is the {\sffamily BINGO} effect prevalent? }

Optimization explores the space of options, looking for good solutions.
The thesis of this paper is that optimization can be made very simple
for problems where those options clump to a very small number of buckets. 
This section reports on how many buckets 
occur in the MOOT data sets of Table~\ref{combinedtable}. After that,
the rest of this paper explores
optimization algorithms that execute over that data.

To understand  this concept of ``buckets'',   consider $d$ \textbf{dimensions} (features) split 
into $b$ \textbf{bins} (ranges). Data rows are assigned to buckets  according to  their
bins per dimension.
For this $b^d$ space,
Figure~\ref{eg1234} offers examples  where  increasing the number of data divisions results
in empty buckets:
\begin{itemize}
\item
EG~\#1 in Figure~\ref{eg1234} shows  that
if   one-dimensional data is spread out, then   more buckets  are filled by  increasing the number of bins from (e.g.) two to four.
\item
On the other hand, as shown in EG~\#2, if the data clumps together, then doubling the number  of bins can lead to  empty buckets.
In this example, increasing the number of bins from two to four leads
to 50\% empty buckets.
\item
Similarly, if two-dimensional  data is spread out, adding dimensions fills more buckets. As in EG~\#3, doubling the dimensions leads to double the number of filled buckets.
\item
On the other hand, as in EG~\#4,  if two-dimensional data clumps, adding new dimensions may yield empty buckets.
\end{itemize}
The lesson of  Figure~\ref{eg1234} is that  
when data clumps, increasing the divisions of the data does
not increase the used parts of those divisions (which is the {\sffamily BINGO} effect). 

\begin{algorithm}[!t]
 \footnotesize
    \caption{Bucket Building with Randomized Parameters}
    \label{bingrows}
    \begin{flushleft}
    \textbf{Input:} Table~\ref{combinedtable} - a list of datasets with rows and columns \\
\hspace*{1em} $\mathit{Buckets(data, d, b)}$ — divide on $b$ bins, $d$ dims, see Table~\ref{bingrow}. \\
\hspace*{1em} $\mathit{RandomInteger(a, b)}$ — returns random integer in $[a,b]$ \\
\hspace*{1em} $\mathit{RandomChoice(Table~1)}$ — selects a random dataset \\
    \textbf{Output:} Two bucket sets $n_1$ and $n_2$ with sufficient row counts 
    \\
    1. $count \gets 0$ \\
    2. \textbf{While} $count < 1000$ \textbf{do} -- try this, many times\\
    3. \hspace*{1em} $d_1 \gets RandomInteger(3..8)$\\
    4. \hspace*{1em} $d_2 \gets RandomInteger(d_1..8)$\\
    4. \hspace*{1em}  $b_1 \gets RandomInteger(3..10)$ \\
    5. \hspace*{1em} $b_2 \gets RandomInteger(b_1..10)$ \\
    6. \hspace*{1em} \textbf{if} $d_2 > d_1$ \textbf{or} $b_2 > b_1$ \textbf{then} -— data divisions have been found\\
    7. \hspace*{2em} $count \gets count + 1$ \\
    8. \hspace*{2em} $data \gets \mathit{RandomChoice}(\text{from Table 1})$ \\
    9. \hspace*{2em} \textbf{for} $(d, b) \in \{(d_1, b_1),~(d_2, b_2)\}$ \textbf{do} \\
    10. \hspace*{3em} $n \gets \{x \in$ buckets$(data, d, b)~|~\text{len}(x) \ge 2d\}$ \\
    11. \hspace*{3em} Append $|n|$ to result list \\
    12. \hspace*{2em} end for\\
    13. \hspace*{2em} \textbf{print} $\left(\lfloor \log_2(|\text{data.rows}|) \rfloor,~\text{result}[0],~\text{result}[1] \right)$ \\
    14. \hspace*{1em} \textbf{end if}\\
    15. \textbf{end while}
    \end{flushleft}
\end{algorithm}
How prevalent is the {\sffamily BINGO} effect in the data of Table~\ref{combinedtable}?
To check that, Algorithm~\ref{bingrows}   counts how many more buckets are filled, after adding more divisions to the data.
That algorithm finds its dimensions using the methods of 
Table~\ref{bingrow}.
To avoid noise and outliers, our buckets must  have at least {\em minPts} number of data rows. Following advice from the DBSCAN community~\cite{Schubert2017},  line 10 of Algorithm~\ref{bingrow},
says   filled buckets need at least $2d$ {\em minPts}.

\begin{table} 
 \caption{Finding Multi-Dimensional Buckets.}\label{bingrow}
\begin{tabular}{|p{.95\linewidth}|}\hline
Data often lies on a simpler, lower-dimensional
"manifold" \cite{zhu2005semi}. To best divide data, dimensions should
reflect that underlying manifold  shape.\\\rowcolor{blue!10} 
 
 Pearson's 1901 {\bf PCA} \cite{pearson1901principal} is a traditional data
reduction method that seeks the overall shape of the data. It identifies  eigenvectors (directions of
max variance) from the data's covariance matrix. These principal
components are ranked by eigenvalues(variance captured). Data
is projected to lower dimensions using the top few eigenvectors.
For large datasets, PCA is costly. \\
The Nyström method
approximates eigendecomposition for large kernel matrices via data
samples, aiding component estimation, notably in Kernel PCA.
 {\bf Fastmap} \cite{faloutsos1995fastmap}, a Nyström-like method, derives
components from distant data "corners" ($k$).
Operation:
\begin{itemize}
    \item Pick random row $r$.
    \item Select row $k_1$ far from $r$.
    \item Select row $k_2$ far from $k_1$.
\end{itemize}
The line $k_1-k_2$, akin to PCA's first principal component,
captures max data spread.\\\rowcolor{blue!10}
Other rows $i$ map to $b$ bins on this
line via the cosine rule:
\begin{equation}\label{fmap}
x=\text{int}\left(\frac{D(k_1,i)^2 + D(k_1,k_2)^2 - D(k_2,i)^2}
{2 \cdot D(k_1,k_2)^2 \cdot b}\right)
\end{equation}
\\
\begin{wrapfigure}{r}{1in}
\includegraphics[width=1in]{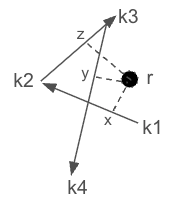}
\end{wrapfigure}
Eq.~\ref{fmap} maps rows to $b$ bins along one dimension defined
by corners $k_1,k_2$. For multiple dimensions, more corners
($k_3, \dots$) are chosen, each maximally distant from prior
corners. Four corners might yield the three dimensions of the figure at right. Rows map to
bins on each dimension (e.g., $r \to (x,y,z)$).
\vspace{10mm}
\\\rowcolor{blue!10}
 For data with numbers and  symbols (e.g., Makefile booleans in
Table~\ref{combinedtable}), Fastmap needs a suitable distance
metric. In {\bf Aha's  distance measure} \cite{aha1991instance}:
\begin{itemize}
    \item Numbers: $\mathit{dist}(x,y) = |x-y|$ (after normalizing numbers 0..1, min..max).
    \item Syms: $\mathit{dist}(x,y) = 0$ if $x=y$, else $1$.
    \item Missing: If $x,y$ missing, $\mathit{dist}(x,y)=1$. If only
    $x$ missing, $x$ is $0$ if $y>0.5$, else $1$ (vice-versa
    for $y$).
\end{itemize}
For vectors: $D(x,y) = \left(\sqrt{\frac{\sum_{i=1}^N \mathit{dist}(x_i,y_i)^2}{N}}\right)/\sqrt{N}$;
$0\le D \le 1$.\\\hline
 \end{tabular}
 \vspace{0.3cm}
 \end{table}
 \begin{figure} 
\caption{ Algorithm~\ref{bingrows}, applied to Table~\ref{combinedtable}. {\em Buckets (after)} counts the  filled buckets seen {\em after}  increasing the data divisions.   Y-axis shows mean results for data sets grouped  by $\log_2$ of the number of rows. Blue lines indicate dataset sizes. Red lines show the bucket counts from randomly chosen bins/dimensions. Yellow lines show buckets after using more bins/dimensions.  }\label{bingos}
\includegraphics[width=3.5in]{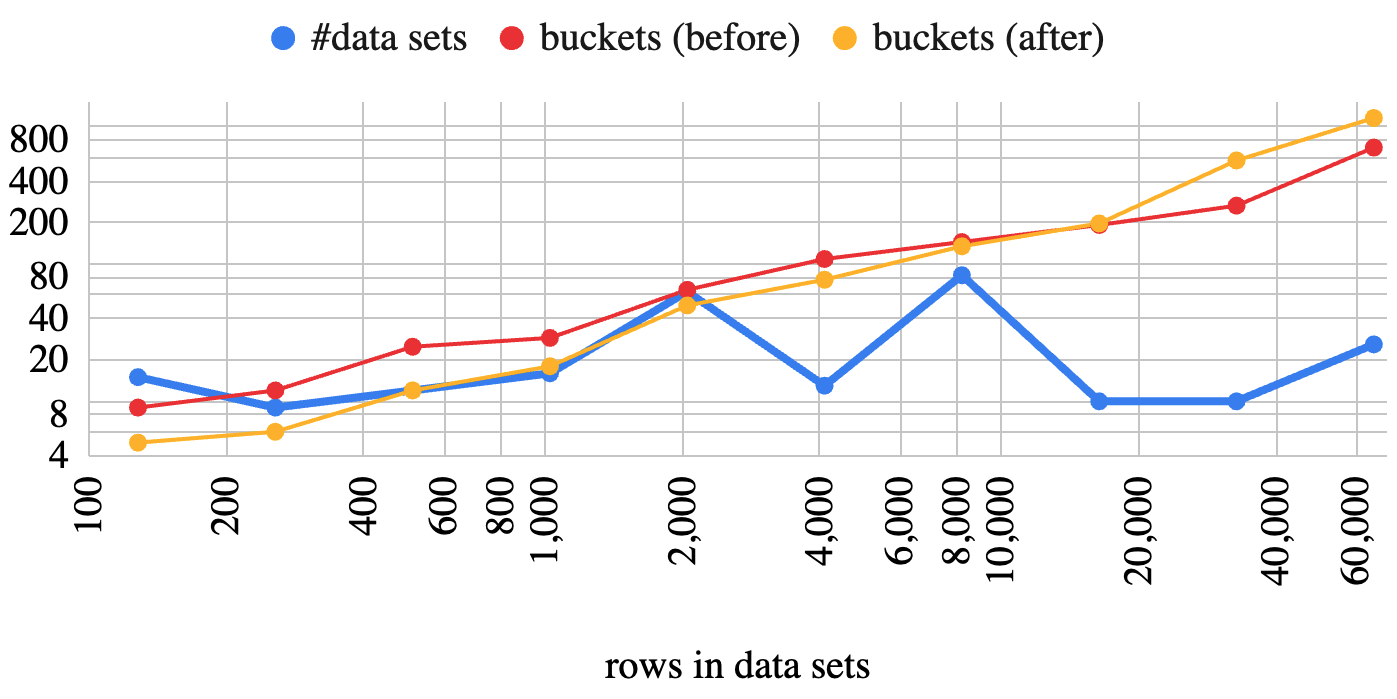}
\end{figure}

Before showing the results, it
is important to   stress the following: 
\begin{quote}
{\em Algorithm~\ref{bingrows} finds buckets using only the   
$x$ values.}
\end{quote}

This may seem like a minor detail but, in practice, it has enormous practical implications. If buckets are generated and counted 
{\em without} using the $y$ dependent values, that means this can happen
{\em without} some expensive and lengthy pre-processing stage to label the data.

  Figure~\ref{bingos} shows the output of Algorithm~\ref{bingrows} applied to   Table~\ref{combinedtable}.  The blue lines shows that most of our data falls in the range 1,000 to 10,000 rows.
The red lines shows the   buckets   generated using some number of   bins and dimensions. The yellow line shows     buckets seen after
trying to use more bins and/or more dimensions.   Figure~\ref{bingos} shows: 
\begin{itemize}
\item
Most buckets are empty (which is the {\sffamily BINGO} effect). Given the ranges of $b$ bins and $d$ dimensions explored in Algorithm~\ref{bingrow},
the mean $b^d$ is around 5000. Yet as shown in this figure, data compresses   to around 100 buckets (see the median $y$ values).     
\item
This data set compression   can improve with size:
\begin{itemize}
\item
 256 rows go to 10 buckets $(10/256\approx 4\%) $
\item  32,000  rows go to 500 buckets
$(500/32,000\approx 1\%)$.
\item  64,000 go to around
1\% of the possible buckets.  
\end{itemize}
\item For most of our samples, the yellow line is below the red line. That is, increasing the number of data divisions does not increase the number of filled buckets.
\end{itemize}
In summary, based on the above:

 \begin{conclusion}{RQ1) Is the {\sffamily BINGO} effect prevalent? }
Empirically, it cannot be shown that this method applies to all data. 
That said, in all the 39 SE optimization
problems of Table I and II, the number of filled buckets is very small.
\end{conclusion}

\section{Methods}

Having answered {\bf RQ1}, we now turn to the other research questions. This section describes the algorithms, evaluation measures, statistical methods, and experimental rig needed to address {\bf RQ2, RQ3, RQ4}.

\subsection{Algorithms}

There are many strategies for optimization. In terms of the {\sffamily BINGO}, 
one important distinction is between
{\em  pool-based} inference and {\em membership query} inference~\cite{settles2012active}. 

The key difference in these
two approaches is  how they treat the filled buckets: pool-based inference stays within the buckets while  membership query may roam elsewhere.

\subsubsection{Membership Query Inference}
In this approach, inference
 can construct arbitrary rows (not necessarily from some pre-defined  pool) and ask for their labels.
 Examples of this kind of inference   
 include evolutionary programs  
 that mutate and combine prior solutions to create new solutions.
 For example, 
DEHB (short for Distributed Evolutionary Hyperband) \cite{awadijcai2021p296}
 is an optimizer  that combines the   search
capabilities of {\em differential evolution} (DE)  with the adaptive
resource allocation of {\em Hyperband}:
\bi
\item
Hyperband \cite{li2018hyperband} runs  by repeatedly
using   {\em successive halving}     across varying
total resource budgets. Successive halving~\cite{jamieson2016nonstochastic}
 runs a set of configurations
with a small budget, discards the worst performers, and promotes the
remaining to the next stage with a larger budget, efficiently focusing
resources on the most promising ones.
\item
Differential Evolution (DE) \cite{storn1997differential}   generates new   solutions by
applying mutation and crossover operations to existing population
vectors, selecting superior   vectors to evolve the population. Unlike standard
genetic algorithms (which mutate bit-by-bit), DE generates mutants by extrapolating between population of known good solutions. 
\item
At runtime,   DE-generated candidates are   evaluated by 
Hyperband, using successive halving to
dynamically allocate resources and swiftly prune poor configurations.
\ei
For the purposes of this paper, the key aspect of DEHB is that it   creates
mutants outside  the initial population of buckets.

DEHB outperforms its predecessor (BOHB~\cite{awadijcai2021p296}) (and BOHB is known  to outperform Hyperopt~\cite{bergstra2015hyperopt}). 
Our reading of the AI literature is that, DEHB~\cite{awadijcai2021p296}  is arguably the current state-of-the-art
in combing evolutionary methods (using DE's differential evolution), multi-fidelity methods (using HB's Hyperband), and iterated racing
(using the sequential halving).

\subsubsection{Pool-based Inference}

Pool-based inference assumes a large pool of unlabeled data from which
the inference selects the most informative examples   to query for labels.
 Examples of pool-based inference include  
 \underline{\em RANDOM}  
 and two of our own algorithms called \underline{\em LINE} and \underline{\em LITE}.

 Given a budget $B$, \underline{\em RANDOM}  selects any $B$ rows, labels and sorts them (see \S\ref{eval}), then returns the best.
Despite its simplicity, RANDOM can yield surprisingly  good results~\cite{bergstra2011algorithms}
(maybe since RANDOM  samples many of the buckets).

\underline{\em LINE}  samples  data points that are different from those already chosen.
Many researchers argue that such {\em diversity sampling} is a good way to explore data~\cite{ijcai2024p540}.     Diversity   sampling methods are often cluster-based \cite{yehuda2018cluster}.
For example, our
LINE algorithm  generates a diverse sample using the centroid finder of the K-Means++ algorithm~\cite{Arthur2007kmeansplusplus}:
\begin{itemize}
\item
Select centroid $c_1$ uniformly at random from dataset $X$. 
\item
Subsequently, for each remaining row $x \in X$, compute its squared distance $D(x)^2 = \min_{c \in C} \|x - c\|^2$ to the closest already chosen centroid $c \in C$. 
\item Then, a new centroid $x'$ is selected with a probability proportional to $D(x')^2$. 
\item
Repeat until $k$ centroids are chosen.
\item
Label all centroids, sort, and return the best one.
\end{itemize}
\underline{\em LITE}  was inspired by  Bergstra's et al.'s  \textit{Tree of Parzen Estimators}~\cite{bergstra2012random}.
TPE is an 
 active learner~\cite{settles2009active}; i.e. it   incrementally  builds and reflects over a {\em surrogate model}
  to decide what row of data to label next.
One way to build the surrogate is  to use   
\textit{Gaussian Process Models}.
These models are a collection of random variables, any finite
number of which have a joint Gaussian distribution \cite{rasmussen2006gaussian}. They define a distribution over functions that can report the mean and variance of any prediction. 
Hence, they can be used by {\em acquisition functions} to  guide the labeling to  rows that (e.g.) need more sampling (since the variance of the labels in that region is very high). 
  
Faster than Gaussian Process Models~\cite{nair2018finding} are surrogates
built via \textit{Tree of Parzen Estimators (TPE)}~\cite{bergstra2012random}.
TPE  divides the labelled data into {\em good} and {\em rest} then models their   distributions   using Parzen estimators
(kernel density estimators). 

LITE was an experiment on simplifying TPE. Instead of Parzen estimators,
it sorts the labelled data into $\sqrt{N}$ {\em good} and  $N-\sqrt{N}$ 
{\em rest}. It then built a two-class classifier that can report the likelihood of an unlabeled example being good $g$ or rest $r$.  LITE
  labels the next row that   maximizes $g/r$ (i.e. that most {\em good} and least {rest}). The updated labeled rows are then sorted and the process loops.
  
  (Aside: Other acquisition functions were explored for LITE that transitioned from exploring high-variance regions
to exploiting regions with good mean scores. 
There are not reported here   since,
for the data of Table~\ref{combinedtable} their behaviors were all statistically indistinguishable).

\subsection{Experimental Rig} 
\subsubsection{Initialization}
All our algorithms begin by taking a data set and randomly shuffling the row order.

After that, LITE and DEHB incrementally modify an initial set of selected rows.
To start that process, LITE  starts with four labeled randomly selected rows.  As to DEHB, it has its initial population, which is unlabeled 
(till the inference starts).

\subsubsection{Labeling}

For pool-based inference tools like LINE, LITE, and RANDOM, labeling is straightforward: it simply involves inspecting the y values of a given row.

However, for membership query inference tools such as DEHB, labeling is more complex since DEHB can generate new rows. This means that DEHB has to   assign values to goals for previously unseen examples. Following the procedure outlined by Pfisterer and Zela et al.~\cite{pfisterer2022yahpo, zela2022}, new rows are labeled by finding their nearest neighbor in Table~\ref{combinedtable}.

Some consideration was given to building a predictive model for labeling previously unseen examples. But it was  realized that the only way to validate such a system would be through the Pfisterer and Zela procedure, as Table~\ref{combinedtable} holds all the  ground truth. Since some form of nearest-neighbor testing is unavoidable,
this paper adopts it from the outset.

\subsubsection{Termination}
All our algorithms run for a fixed budget (number of rows to sample).
Initialization (described above) consumes some init number of that budget.

Recalling  \S\ref{hardlabel},
the budgets  are  initialized to something   small (i.e. six samples), then increased   while those increases lead to better performance.

\subsection{Evaluation Measures}\label{eval}
As mentioned above, one of our primary evaluation criteria 
is   minimizing the {\em cost}  required to {\em benefit} from finding good rows in
the MOOT data sets. {\em Cost} will be measured in two ways:
\begin{itemize}
\item Number of accessed labels.
\item Algorithm runtimes. For the sake of uniformity, all
codes are run on the same hardware\footnote{A standard Linux laptop with 4-core 1.30 GHz CPU, 16GB of RAM, and no GPU.}.
\end{itemize}
But how to measure {\em benefit}? Many of our data sets are multi-objective which means a good result for one goal might be a bad result
for another.

This study uses ``distance to heaven''.
For dependent attributes that are to be  minimized or maximized, the ideal   $y$-values contain the minimum or maximum value (respectively) for that column. 
For example, in Figure~\ref{moot}, the $y$ columns are labelled 
{\em Throughput+, Latency-}; i.e. we want to maximize the first and minimize the second.
After normalizing the dependent values to a 0..1 range (min to max), the ideal point - called {\em heaven} -is $\{1,0\}$. Any row in that data set can be scored on how far away it falls from that point.  Note that, for this measure,
{\em lower} values are {\em better},
  since {\em smaller} values are {\em closer} to the ideal\footnote{
 There are many other ways to measure {\em benefit} in multi-objective reasoning.
A recent IEEE TSE article by Chen et al. \cite{Li22} reviewed various multi-objective optimization performance measures like Hypervolume (HV), Spread ($\Delta$), Generational Distance (GD), and Inverted Generational Distance (IGD). We consulted with
those authors on this paper and they offered the opinion~\cite{doe2023personal} that their measures are inappropriate
this paper
due to our focus on minimal labeling.
For instance, our approach would inherently result in low hypervolume since we generate very few solutions.}.

In order to get a quick overview of all our results, the following summary statistic is defined. For every dataset in Table~\ref{combinedtable},
there exists a mean $\mu$ and a {\em min} distance to heaven. After each run, our  optimizers return a row with a distance to heaven score of $x$. After 20 repeated runs,
these optimizers return best rows
with a mean score of $x$. 
We define:

\begin{table*}[t]
\centering
\scriptsize
\caption{Optimization performance over the Table~\ref{combinedtable} data.
  Numbers are means seen in 20 repeats.
The $\Delta$ of column one comes from Equation~\ref{delta} (larger numbers are better).
  In the results table, cell background colors indicate come from the statistical
  measures of \S\ref{stats} and show the relative performance rankings:
\raisebox{0.5ex}{\protect\fcolorbox{black}{white}{\rule{1em}{0em}}} best,
\raisebox{0.5ex}{\protect\fcolorbox{black}{lightyellow}{\rule{1em}{0em}}} second-best,
\raisebox{0.5ex}{\protect\fcolorbox{black}{lightpurple}{\rule{1em}{0em}}} third-best,
\raisebox{0.5ex}{\protect\fcolorbox{black}{lightred}{\rule{1em}{0em}}} fourth-best,
and \raisebox{0.5ex}{\protect\fcolorbox{black}{lightgrey}{\rule{1em}{0em}}} fifth-best results. On the last row,
the black cells
\raisebox{0.5ex}{\protect\fcolorbox{black}{black}{\rule{1em}{0em}}}
show the percentage of that column of treatments which   achieved best results.  Finally, the
green cell
\raisebox{0.5ex}{\protect\fcolorbox{black}{green}{\rule{1em}{0em}}} shows the median optimization result.}
\begin{adjustbox}{width=\linewidth}
{\setlength{\tabcolsep}{2.5pt}
\begin{tabular}{c r r c c | r r r r r r r |r r r r r r r |r r r r r r r |
            r r r r r r r |l}
 
\multicolumn{5}{c}{ ~}   & \multicolumn{7}{c}{DEHB} & \multicolumn{7}{c}{LITE} & \multicolumn{7}{c}{LINE} & \multicolumn{7}{c}{Random} &   \\
\cmidrule(lr){6-12} \cmidrule(lr){13-19} \cmidrule(lr){20-26} \cmidrule(lr){27-33}

$\Delta$ &Rows &$|X|$ & $|Y|$&As is & 6 & 12 & 18 & 24 & 50 & 100 & 200 & 6 & 12 & 18 & 24 & 50 & 100 & 200 & 6 & 12 & 18 & 24 & 50 & 100 & 200 & 6 & 12 & 18 & 24 & 50 & 100 & 200 & \\
\cline{1-33}
49 & 10000 & 5 & 3 & \cellcolor{lightpurple}59 & 43 & 42 & 41 & 41 & 40 & 38 & 36 & 43 & 37 & 35 & 35 & 34 & 31 & 28 & \cellcolor{lightyellow}44 & 41 & 39 & 38 & 36 & 35 & 32 & 42 & 42 & 42 & 41 & 39 & 36 & 34 & Health-hard\\
61 & 86058 & 11 & 2 & \cellcolor{lightred}58 & \cellcolor{lightpurple}42 & \cellcolor{lightyellow}40 & \cellcolor{lightyellow}37 & \cellcolor{lightyellow}34 & 28 & 24 & 20 & \cellcolor{lightyellow}37 & \cellcolor{lightyellow}31 & 27 & 25 & 20 & 16 & 14 & \cellcolor{lightyellow}38 & \cellcolor{lightyellow}34 & 25 & 26 & 24 & 18 & 19 & \cellcolor{lightyellow}40 & \cellcolor{lightyellow}33 & \cellcolor{lightyellow}31 & \cellcolor{lightyellow}30 & 22 & 21 & 20 & SS-X\\
63 & 53662 & 17 & 2 & \cellcolor{lightred}25 & \cellcolor{lightyellow}17 & \cellcolor{lightyellow}16 & \cellcolor{lightyellow}15 & \cellcolor{lightyellow}14 & \cellcolor{lightyellow}12 & 11 & 10 & \cellcolor{lightpurple}19 & \cellcolor{lightyellow}17 & \cellcolor{lightyellow}17 & \cellcolor{lightyellow}16 & \cellcolor{lightyellow}16 & \cellcolor{lightyellow}16 & \cellcolor{lightyellow}16 & \cellcolor{lightyellow}16 & \cellcolor{lightyellow}15 & \cellcolor{lightyellow}14 & \cellcolor{lightyellow}14 & 10 & 8 & 8 & \cellcolor{lightpurple}18 & \cellcolor{lightyellow}16 & \cellcolor{lightyellow}16 & \cellcolor{lightyellow}14 & \cellcolor{lightyellow}14 & \cellcolor{lightyellow}12 & 8 & SS-N\\
72 & 4653 & 38 & 1 & \cellcolor{lightpurple}46 & \cellcolor{lightyellow}27 & \cellcolor{lightyellow}25 & \cellcolor{lightyellow}22 & \cellcolor{lightyellow}20 & 15 & 13 & 12 & \cellcolor{lightyellow}23 & 16 & 14 & 13 & 10 & 8 & 7 & \cellcolor{lightyellow}26 & \cellcolor{lightyellow}20 & 15 & 15 & 14 & 11 & 10 & \cellcolor{lightyellow}23 & \cellcolor{lightyellow}21 & 16 & 17 & 13 & 12 & 9 & SQL\_AllMeasurements\\
76 & 1512 & 3 & 2 & \cellcolor{lightyellow}32 & 8 & 8 & 7 & 6 & 5 & 4 & 4 & 13 & 10 & 9 & 8 & 7 & 6 & 4 & 13 & 9 & 8 & 8 & 6 & 6 & 5 & 13 & 9 & 7 & 6 & 5 & 6 & 5 & SS-C\\
78 & 3008 & 14 & 2 & \cellcolor{lightyellow}41 & 22 & 21 & 19 & 19 & 17 & 16 & 14 & 23 & 21 & 19 & 19 & 17 & 14 & 13 & 23 & 19 & 19 & 18 & 17 & 16 & 16 & 23 & 19 & 18 & 19 & 17 & 17 & 16 & SS-R\\
81 & 756 & 3 & 2 & \cellcolor{lightred}41 & \cellcolor{lightyellow}28 & \cellcolor{lightyellow}26 & \cellcolor{lightyellow}24 & \cellcolor{lightyellow}23 & 19 & 17 & 14 & \cellcolor{lightpurple}32 & \cellcolor{lightyellow}28 & \cellcolor{lightyellow}26 & \cellcolor{lightyellow}24 & 21 & 19 & 14 & \cellcolor{lightyellow}24 & 21 & 21 & 22 & 16 & 14 & 13 & \cellcolor{lightyellow}26 & \cellcolor{lightyellow}24 & 21 & 19 & 17 & 14 & 14 & SS-E\\
81 & 1343 & 3 & 2 & \cellcolor{lightyellow}25 & 10 & 8 & 7 & 6 & 3 & 2 & 1 & 8 & 6 & 3 & 2 & 2 & 1 & 0 & 10 & 6 & 6 & 5 & 3 & 3 & 2 & 10 & 7 & 5 & 5 & 4 & 3 & 2 & SS-A\\
81 & 972 & 11 & 2 & \cellcolor{lightpurple}39 & \cellcolor{lightyellow}28 & \cellcolor{lightyellow}27 & \cellcolor{lightyellow}24 & \cellcolor{lightyellow}23 & 21 & 17 & 15 & 21 & 17 & 17 & 17 & 16 & 13 & 13 & \cellcolor{lightyellow}24 & 21 & 20 & 18 & 16 & 15 & 13 & \cellcolor{lightyellow}27 & \cellcolor{lightyellow}22 & 21 & 20 & 17 & 15 & 13 & SS-O\\
81 & 10000 & 88 & 3 & \cellcolor{lightred}52 & \cellcolor{lightpurple}46 & \cellcolor{lightpurple}45 & \cellcolor{lightyellow}43 & \cellcolor{lightyellow}42 & 39 & 37 & 35 & \cellcolor{lightpurple}44 & \cellcolor{lightyellow}40 & 39 & 38 & 36 & 35 & 35 & \cellcolor{lightyellow}43 & \cellcolor{lightyellow}41 & \cellcolor{lightyellow}41 & \cellcolor{lightyellow}40 & 38 & 37 & 37 & \cellcolor{lightyellow}44 & \cellcolor{lightyellow}41 & \cellcolor{lightyellow}42 & \cellcolor{lightyellow}41 & 39 & 38 & 37 & billing10k\\
83 & 6840 & 16 & 2 & \cellcolor{lightpurple}55 & \cellcolor{lightyellow}20 & \cellcolor{lightyellow}18 & 13 & 12 & 10 & 7 & 6 & \cellcolor{lightyellow}24 & 15 & 13 & 10 & 7 & 2 & 2 & \cellcolor{lightyellow}29 & \cellcolor{lightyellow}20 & 15 & 14 & 10 & 4 & 3 & \cellcolor{lightyellow}27 & \cellcolor{lightyellow}19 & 14 & 11 & 10 & 7 & 4 & SS-V\\
90 & 196 & 3 & 2 & \cellcolor{lightred}57 & \cellcolor{lightyellow}45 & \cellcolor{lightyellow}43 & \cellcolor{lightyellow}41 & \cellcolor{lightyellow}40 & 38 & 36 & 35 & \cellcolor{lightpurple}47 & \cellcolor{lightyellow}44 & \cellcolor{lightyellow}43 & \cellcolor{lightyellow}43 & \cellcolor{lightyellow}40 & 35 & 33 & \cellcolor{lightyellow}43 & \cellcolor{lightyellow}42 & \cellcolor{lightyellow}42 & \cellcolor{lightyellow}40 & 37 & 36 & 33 & \cellcolor{lightyellow}44 & \cellcolor{lightyellow}42 & \cellcolor{lightyellow}41 & 39 & 37 & 36 & 34 & SS-G\\
90 & 398 & 4 & 3 & \cellcolor{lightred}56 & \cellcolor{lightpurple}39 & \cellcolor{lightyellow}38 & \cellcolor{lightyellow}35 & \cellcolor{lightyellow}33 & \cellcolor{lightyellow}29 & \cellcolor{lightyellow}27 & 21 & \cellcolor{lightyellow}32 & 25 & 21 & 20 & 18 & 17 & 17 & \cellcolor{lightyellow}36 & \cellcolor{lightyellow}33 & \cellcolor{lightyellow}29 & \cellcolor{lightyellow}29 & 23 & 21 & 18 & \cellcolor{lightyellow}36 & \cellcolor{lightyellow}32 & \cellcolor{lightyellow}32 & 25 & 24 & 20 & 19 & auto93\\
90 & 192 & 9 & 1 & \cellcolor{lightyellow}33 & 11 & 9 & 9 & 6 & 4 & 2 & 2 & 5 & 2 & 1 & 0 & 0 & 0 & 0 & 7 & 4 & 4 & 3 & 1 & 0 & 0 & 9 & 5 & 3 & 3 & 1 & 1 & 0 & Apache\_AllMeasurements\\
90 & 2880 & 6 & 2 & \cellcolor{lightgrey}52 & \cellcolor{lightpurple}30 & \cellcolor{lightpurple}26 & \cellcolor{lightpurple}23 & \cellcolor{lightyellow}20 & \cellcolor{lightyellow}15 & 9 & 5 & \cellcolor{lightred}37 & \cellcolor{lightyellow}19 & \cellcolor{lightyellow}16 & \cellcolor{lightyellow}13 & 8 & 6 & 5 & \cellcolor{lightpurple}31 & \cellcolor{lightyellow}17 & \cellcolor{lightyellow}19 & \cellcolor{lightyellow}15 & 9 & 7 & 4 & \cellcolor{lightpurple}29 & \cellcolor{lightpurple}23 & \cellcolor{lightyellow}19 & \cellcolor{lightyellow}18 & \cellcolor{lightyellow}13 & 7 & 5 & SS-K\\
90 & 3840 & 6 & 2 & \cellcolor{lightred}52 & \cellcolor{lightyellow}28 & \cellcolor{lightyellow}22 & \cellcolor{lightyellow}16 & 13 & 6 & 5 & 4 & \cellcolor{lightpurple}41 & 14 & 11 & 9 & 7 & 6 & 4 & \cellcolor{lightyellow}26 & 11 & 9 & 7 & 5 & 4 & 4 & \cellcolor{lightyellow}22 & \cellcolor{lightyellow}16 & 10 & 8 & 6 & 5 & 4 & SS-J\\
90 & 4608 & 21 & 2 & \cellcolor{lightpurple}46 & 29 & 26 & 25 & 25 & 24 & 23 & 23 & \cellcolor{lightyellow}32 & 30 & 28 & 27 & 24 & 23 & 23 & 29 & 27 & 25 & 25 & 25 & 24 & 24 & 23 & 26 & 25 & 25 & 24 & 24 & 23 & SS-U\\
91 & 196 & 3 & 2 & \cellcolor{lightred}55 & \cellcolor{lightpurple}46 & \cellcolor{lightpurple}44 & \cellcolor{lightyellow}43 & \cellcolor{lightyellow}41 & \cellcolor{lightyellow}38 & 34 & 33 & \cellcolor{lightpurple}46 & \cellcolor{lightyellow}40 & \cellcolor{lightyellow}39 & \cellcolor{lightyellow}37 & 33 & 33 & 32 & \cellcolor{lightyellow}42 & \cellcolor{lightyellow}38 & \cellcolor{lightyellow}38 & 36 & 33 & 32 & 32 & \cellcolor{lightyellow}41 & \cellcolor{lightyellow}40 & \cellcolor{lightyellow}40 & \cellcolor{lightyellow}38 & 36 & 33 & 32 & SS-D\\
\cellcolor{green}91 & 196 & 3 & 2 & \cellcolor{lightred}56 & \cellcolor{lightyellow}44 & \cellcolor{lightyellow}43 & \cellcolor{lightyellow}41 & \cellcolor{lightyellow}41 & 37 & 36 & 35 & \cellcolor{lightpurple}47 & \cellcolor{lightyellow}45 & \cellcolor{lightyellow}44 & \cellcolor{lightyellow}44 & \cellcolor{lightyellow}40 & 36 & 35 & \cellcolor{lightyellow}44 & \cellcolor{lightyellow}44 & \cellcolor{lightyellow}43 & \cellcolor{lightyellow}43 & \cellcolor{lightyellow}41 & 38 & 36 & 35 & \cellcolor{lightyellow}43 & \cellcolor{lightyellow}41 & 39 & 38 & 36 & 35 & SS-F\\
91 & 1080 & 5 & 2 & \cellcolor{lightred}39 & \cellcolor{lightpurple}24 & \cellcolor{lightyellow}18 & \cellcolor{lightyellow}16 & \cellcolor{lightyellow}14 & 7 & 2 & 0 & \cellcolor{lightpurple}23 & \cellcolor{lightyellow}18 & \cellcolor{lightyellow}16 & \cellcolor{lightyellow}13 & \cellcolor{lightyellow}9 & 8 & 2 & \cellcolor{lightyellow}18 & \cellcolor{lightyellow}14 & \cellcolor{lightyellow}13 & \cellcolor{lightyellow}12 & 5 & 2 & 1 & \cellcolor{lightpurple}24 & \cellcolor{lightyellow}18 & \cellcolor{lightyellow}14 & \cellcolor{lightyellow}14 & 6 & 5 & 1 & SS-I\\
92 & 3840 & 6 & 2 & \cellcolor{lightyellow}24 & 4 & 3 & 3 & 2 & 1 & 1 & 0 & 5 & 4 & 3 & 2 & 2 & 2 & 2 & 5 & 2 & 3 & 2 & 2 & 0 & 0 & 4 & 3 & 2 & 1 & 1 & 1 & 0 & SS-S\\
92 & 2866 & 6 & 1 & \cellcolor{lightyellow}23 & 5 & 2 & 2 & 2 & 2 & 1 & 1 & 5 & 2 & 1 & 1 & 1 & 1 & 1 & 2 & 2 & 2 & 1 & 1 & 1 & 1 & 3 & 2 & 2 & 2 & 1 & 1 & 1 & sol-6-c2-obj1\\
92 & 5184 & 12 & 2 & \cellcolor{lightyellow}29 & 8 & 6 & 6 & 6 & 5 & 4 & 4 & 9 & 7 & 7 & 6 & 5 & 5 & 4 & 8 & 5 & 4 & 4 & 4 & 3 & 3 & 7 & 6 & 5 & 5 & 4 & 3 & 3 & SS-T\\
92 & 2736 & 13 & 3 & \cellcolor{lightyellow}41 & 13 & 11 & 9 & 9 & 6 & 5 & 5 & 13 & 6 & 5 & 5 & 5 & 5 & 5 & 11 & 9 & 6 & 7 & 6 & 5 & 5 & 12 & 10 & 8 & 7 & 6 & 5 & 5 & SS-Q\\
92 & 1023 & 11 & 2 & \cellcolor{lightgrey}63 & 25 & 24 & 20 & 17 & 16 & 16 & 16 & \cellcolor{lightpurple}35 & \cellcolor{lightyellow}25 & 22 & 20 & 19 & 17 & 16 & \cellcolor{lightyellow}31 & \cellcolor{lightpurple}35 & \cellcolor{lightyellow}25 & 25 & 19 & 17 & 17 & \cellcolor{lightred}42 & \cellcolor{lightyellow}25 & 23 & 22 & 18 & 17 & 16 & SS-L\\
92 & 65536 & 16 & 2 & \cellcolor{lightgrey}52 & \cellcolor{lightred}34 & \cellcolor{lightred}32 & \cellcolor{lightred}28 & \cellcolor{lightpurple}26 & \cellcolor{lightpurple}22 & \cellcolor{lightyellow}18 & 12 & \cellcolor{lightgrey}39 & \cellcolor{lightred}33 & \cellcolor{lightred}31 & \cellcolor{lightred}30 & \cellcolor{lightpurple}23 & \cellcolor{lightyellow}20 & \cellcolor{lightyellow}17 & \cellcolor{lightgrey}39 & \cellcolor{lightred}33 & \cellcolor{lightred}33 & \cellcolor{lightred}33 & \cellcolor{lightred}29 & \cellcolor{lightpurple}25 & \cellcolor{lightpurple}23 & \cellcolor{lightgrey}38 & \cellcolor{lightred}33 & \cellcolor{lightred}33 & \cellcolor{lightred}31 & \cellcolor{lightred}29 & \cellcolor{lightpurple}27 & \cellcolor{lightpurple}24 & SS-W\\
93 & 206 & 3 & 2 & \cellcolor{lightred}56 & \cellcolor{lightpurple}32 & \cellcolor{lightpurple}30 & \cellcolor{lightpurple}27 & \cellcolor{lightpurple}24 & \cellcolor{lightyellow}17 & \cellcolor{lightyellow}11 & 6 & \cellcolor{lightpurple}25 & 9 & 4 & 3 & 3 & 1 & 0 & \cellcolor{lightpurple}27 & \cellcolor{lightyellow}21 & \cellcolor{lightyellow}21 & \cellcolor{lightyellow}16 & \cellcolor{lightyellow}11 & 9 & 0 & \cellcolor{lightpurple}30 & \cellcolor{lightyellow}23 & \cellcolor{lightyellow}20 & \cellcolor{lightyellow}16 & \cellcolor{lightyellow}15 & 4 & 3 & SS-B\\
94 & 196 & 3 & 1 & \cellcolor{lightpurple}27 & 7 & 4 & 3 & 3 & 0 & 0 & 0 & 6 & 1 & 0 & 0 & 0 & 0 & 0 & 4 & 2 & 2 & 1 & 0 & 0 & 0 & \cellcolor{lightyellow}10 & 3 & 1 & 0 & 0 & 0 & 0 & wc+wc-3-c4-obj1\\
94 & 3456 & 14 & 1 & \cellcolor{lightyellow}12 & 2 & 1 & 1 & 1 & 0 & 0 & 0 & 2 & 1 & 1 & 0 & 0 & 0 & 0 & 2 & 1 & 1 & 1 & 0 & 0 & 0 & 2 & 1 & 1 & 1 & 0 & 0 & 0 & HSMGP\_num\\
95 & 196 & 3 & 1 & \cellcolor{lightpurple}26 & 6 & 3 & 2 & 2 & 1 & 0 & 0 & 6 & 1 & 0 & 0 & 0 & 0 & 0 & 4 & 2 & 2 & 1 & 0 & 0 & 0 & \cellcolor{lightyellow}9 & 3 & 0 & 0 & 0 & 0 & 0 & wc+sol-3-c4-obj1\\
95 & 196 & 3 & 1 & \cellcolor{lightpurple}27 & 6 & 4 & 2 & 1 & 1 & 0 & 0 & 6 & 1 & 0 & 0 & 0 & 0 & 0 & 4 & 3 & 2 & 1 & 0 & 0 & 0 & \cellcolor{lightyellow}9 & 3 & 1 & 0 & 0 & 0 & 0 & wc+rs-3-c4-obj1\\
95 & 259 & 4 & 2 & \cellcolor{lightred}71 & \cellcolor{lightpurple}47 & \cellcolor{lightyellow}40 & \cellcolor{lightyellow}35 & \cellcolor{lightyellow}33 & 28 & 24 & 23 & \cellcolor{lightpurple}43 & 26 & 25 & 25 & 24 & 23 & 23 & \cellcolor{lightpurple}49 & \cellcolor{lightyellow}38 & \cellcolor{lightyellow}36 & \cellcolor{lightyellow}36 & 30 & 26 & 24 & \cellcolor{lightpurple}43 & \cellcolor{lightyellow}41 & \cellcolor{lightyellow}35 & \cellcolor{lightyellow}35 & 27 & 23 & 23 & SS-H\\
95 & 1023 & 11 & 2 & \cellcolor{lightred}63 & \cellcolor{lightyellow}25 & \cellcolor{lightyellow}24 & 20 & 17 & 16 & 16 & 16 & \cellcolor{lightpurple}35 & 23 & 20 & 16 & 16 & 16 & 16 & \cellcolor{lightpurple}38 & \cellcolor{lightyellow}29 & \cellcolor{lightyellow}24 & \cellcolor{lightyellow}26 & 18 & 17 & 16 & \cellcolor{lightpurple}42 & \cellcolor{lightyellow}25 & 23 & 22 & 18 & 17 & 16 & SS-P\\
95 & 2880 & 6 & 1 & \cellcolor{lightpurple}29 & 7 & 5 & 4 & 1 & 1 & 0 & 0 & \cellcolor{lightyellow}10 & 2 & 0 & 0 & 0 & 0 & 0 & 6 & 2 & 1 & 1 & 0 & 0 & 0 & 5 & 2 & 1 & 1 & 0 & 0 & 0 & wc-6-c1-obj1\\
95 & 1152 & 16 & 1 & \cellcolor{lightyellow}38 & 3 & 2 & 1 & 1 & 0 & 0 & 0 & 7 & 3 & 0 & 0 & 0 & 0 & 0 & 7 & 3 & 2 & 1 & 1 & 0 & 0 & 8 & 3 & 2 & 2 & 1 & 0 & 0 & X264\_AllMeasurements\\
96 & 10000 & 5 & 3 & \cellcolor{lightpurple}66 & \cellcolor{lightyellow}29 & 16 & 8 & 4 & 4 & 0 & 0 & 13 & 0 & 0 & 0 & 0 & 0 & 0 & 14 & 0 & 0 & 0 & 0 & 0 & 0 & \cellcolor{lightyellow}28 & 4 & 4 & 0 & 0 & 0 & 0 & Health-easy\\
96 & 864 & 17 & 3 & \cellcolor{lightyellow}31 & 9 & 4 & 3 & 2 & 1 & 0 & 0 & 3 & 2 & 1 & 1 & 1 & 0 & 0 & 3 & 1 & 1 & 1 & 0 & 0 & 0 & 5 & 2 & 1 & 1 & 0 & 0 & 0 & SS-M\\
97 & 3840 & 6 & 1 & \cellcolor{lightyellow}28 & 2 & 1 & 0 & 0 & 0 & 0 & 0 & 4 & 0 & 0 & 0 & 0 & 0 & 0 & 2 & 1 & 1 & 0 & 0 & 0 & 0 & 4 & 1 & 0 & 0 & 0 & 0 & 0 & rs-6-c3\_obj1\\
100 & 3840 & 6 & 1 & \cellcolor{lightyellow}7 & 0 & 0 & 0 & 0 & 0 & 0 & 0 & 0 & 0 & 0 & 0 & 0 & 0 & 0 & 0 & 0 & 0 & 0 & 0 & 0 & 0 & 0 & 0 & 0 & 0 & 0 & 0 & 0 & rs-6-c3\_obj2\\
\rowcolor{lightblack}
\multicolumn{4}{|c|}{\textcolor{white}{\%Best}} & \textcolor{white}{0} & \textcolor{white}{51} & \textcolor{white}{54} & \textcolor{white}{59} & \textcolor{white}{62} & \textcolor{white}{85} & \textcolor{white}{92} & \textcolor{white}{100} & \textcolor{white}{49} & \textcolor{white}{72} & \textcolor{white}{79} & \textcolor{white}{79} & \textcolor{white}{87} & \textcolor{white}{95} & \textcolor{white}{95} & \textcolor{white}{49} & \textcolor{white}{59} & \textcolor{white}{67} & \textcolor{white}{72} & \textcolor{white}{95} & \textcolor{white}{97} & \textcolor{white}{97} & \textcolor{white}{41} & \textcolor{white}{51} & \textcolor{white}{69} & \textcolor{white}{77} & \textcolor{white}{90} & \textcolor{white}{95} & \textcolor{white}{97}\\
\end{tabular}}
\end{adjustbox}

\label{tab:results}
\end{table*}

\begin{equation}\label{delta}
\Delta = 100 * \left(1 - \frac{x-\mathit{min}}{\mu - \mathit{min}} \right)
\end{equation}
For our methods,   $0 \le \Delta \le 100$ and
\bi
\item
$\Delta = 0$ is a failure since, on average, optimization does not improve  initial conditions;
\item
  $\Delta=100$ is a    success since, on average,  optimizers  find the best
row with {\em min} distance to heaven.
\ei


\subsection{Statistical Methods}\label{stats}
Our algorithms are stochastic:
\bi
\item Initially, all the algorithms shuffle the row order;
\item DEHB explores random mutations;
\item RANDOM selects rows stochastically;
\item LINE picks new centroids at random according to their distances from existing centroids;
\item LITE starts its work by picking four rows at random.
\ei
Accordingly, all our results are repeated 20 times, each time with different random number seeds. Different treatments are then ranked via   statistical methods.

   Scott-Knott \cite{Scott1974} is a
recursive bi-clustering statistical method. Results from different sample sizes
and algorithms are collected from 20 repeated runs.
These treatments are then sorted by their median distance to heaven.  Scott-Knott splits the
list where the expected difference in means is largest:

{\scriptsize \begin{equation}
E(\Delta) = \frac{|l_1|}{l} abs(E(l_1)-E(l))^2 + 
\frac{|l_2|}{l} abs(E(l_2)-E(l))^2
\end{equation}}

Here, $|l_1|$ and $|l_2|$ are the sizes of the split lists.
If an effect size test and a significance test   confirm differences,
Scott-Knott ranks the groups and recurses on each split.
The leaves of the resulting tree are then ranked from left to right as 1,2,3...
Treatments are ranked according to their leaf in the 
  tree.

For an effect size test, these statistics use a non-parametric effect size metric
call  Cliff's Delta~\cite{macbeth2011cliff}.
This measures how often values from one group exceed another.
The score ranges from -1 (entirely less) to 1 (entirely more),
with 0 meaning no difference. It needs no normality
assumptions and classifies effects as small, medium, or large. Scott-Knott only recurses for medium to large effects. 

For the significance test, these statistics use the {\em Bootstrapping} is a resampling method~\cite{EfroTibs93}. It draws samples
with replacement to build distributions of stats (mean, median)
and estimate confidence intervals. It tests if results differ
from chance, useful with small or skewed datasets.

Scott-Knott is preferred over post-hoc tests like Nemenyi \cite{nemenyi1963distribution}
since it avoids overlapping groups. It tests both significance
and effect size. It
handles overlapping distributions well, needing only
$\log_2(N)$ comparisons for N samples.

\section{Results}\label{results}
Table~\ref{tab:results} shows results from 20  trials, using   budgets:
\vspace{-0.5em}
\[
\mathit{budget}=\{6,12,18,24,50,100,200\}  \vspace{-0.5em}\]

We stop sampling at a budget of 200 since, after 100 samples, we rarely found anything ``better''`.

(Note: we say that one cell is ``better'' than another if it has a ``better-ranking'' color.
For each row, the white, yellow, purple, pink and gray cells denote 
results ranked best, second,-best third, fourth, and fifth by the statistical methods of \S\ref{stats}.)

\begin{figure} 
\begin{center} \includegraphics[width=3.05in]{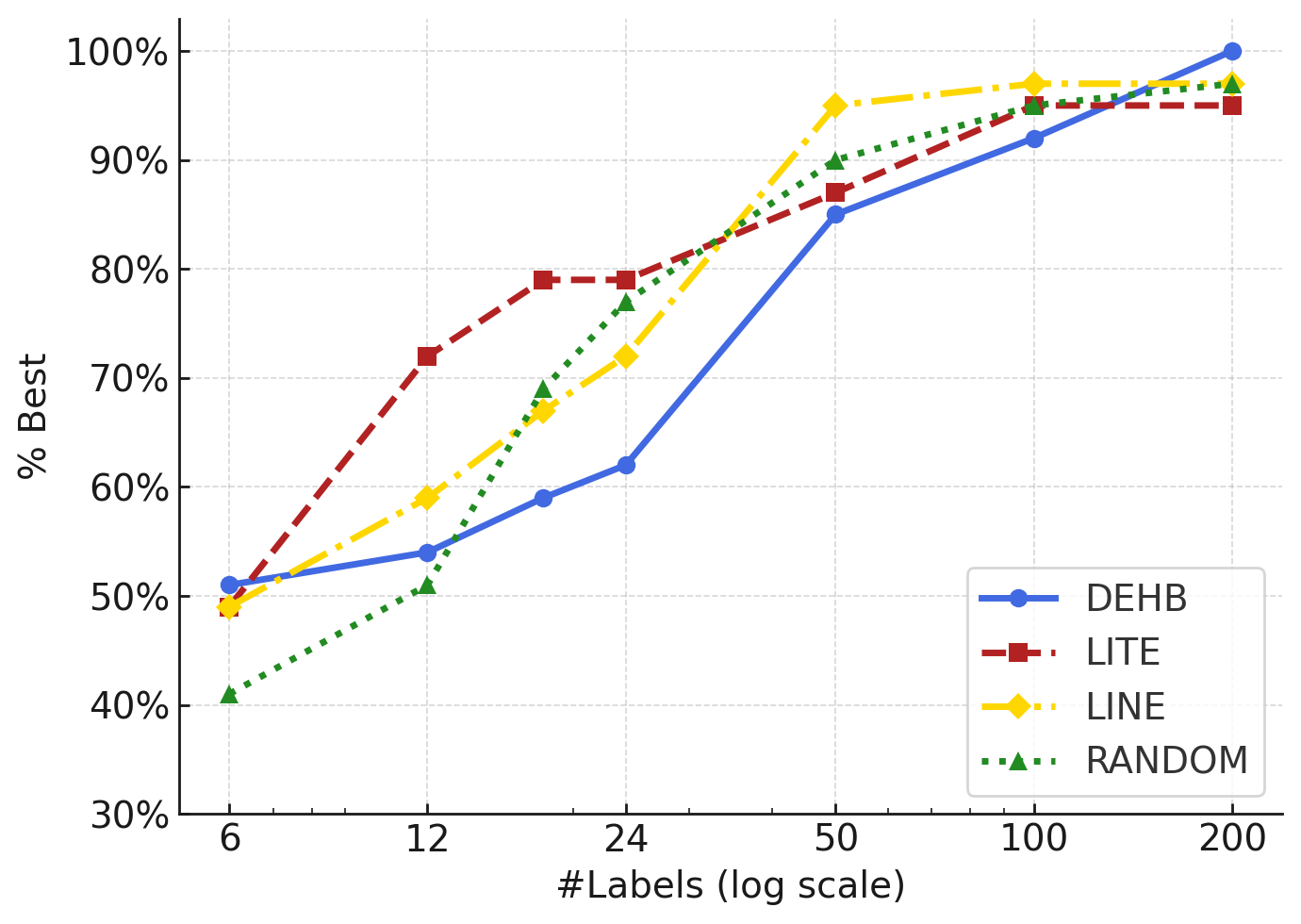}
\end{center}
\caption{For each sample size, in what percentage of the data sets, did a treatment achieve best ranked performance? Data from last row of  Table~\ref{tab:results}. }\label{pcf}
\vspace{-5mm}
\end{figure}

The {\em As is} column of Table~\ref{tab:results} shows  the $\mu$ distances to heaven of the data, before optimization. In all data sets,  to the right of {\em As is},   there is always one or more better cells; i.e.
\begin{quote}
{\em Our optimizers find     significant    improvements.}
\end{quote}
The 
rows  of Table~\ref{tab:results} are sorted by    Equation~\ref{delta}.
Table~\ref{tab:results}'s  green cell lists   
 the median optimization: $\Delta=91\%$; i.e. 
 \begin{quote}
{\em Our optimizers find   very large improvements.}
\end{quote}
 We tried, unsuccessfully,  to learn what kind of data sets were ranked above or below the median.  
 After applying decision trees, regression, and clustering, we   concluded:
\begin{quote}
{\em  Static attributes of these data sets
(i.e. {\em rows}, $|x|$, $|y|$)  do
not predict optimizer performance.}
\end{quote}
The black cells in the last row of
Table~\ref{tab:results} show how often (in percentages) an optimizer achieved first-ranked results. 
Figure~\ref{pcf} summarizes those numbers.
In that figure, we see that DEHB usually performs worse than anything else. This is significant since DEHB was our membership query inference method that could think ``outside of the buckets''. Hence we say that, at least for these data sets:

\begin{conclusion}{ RQ2) Is   BINGO-based reasoning effective?}
 BINGO-based reasoning
(that restricts itself to the filled buckets) works as well as    methods that explore elsewhere.   
\end{conclusion}
(To say the least, this observation challenges decades 
of research that recommends mutation-based methods.)  

Different optimizers work best for different sample sizes:
 \begin{quote}
{\em For labels $\le 24$, LITE's  Bayesian sampling   is  best. }
\end{quote}
Also:
 \begin{quote}
{\em  For
  labels $> 24$, LINE's diversity sampling  is best }
\end{quote}
Above 50 samples, many of our methods start to plateau. For mission-critical and safety-critical applications, there would be benefit in exploring more samples. But for many engineering applications, achieving over 90\% of the optimal after just 50 labels would be very useful. 

As to that plateau effect after 50 samples,  we conjecture: 
\begin{quote}
{\em  There could be a double BINGO effect; i.e. data   falls to just a few buckets, of which only  some small subset  needs to be explored.} 
\end{quote}
\begin{figure*}[t]
        \centering
        \includegraphics[width=.7\linewidth]{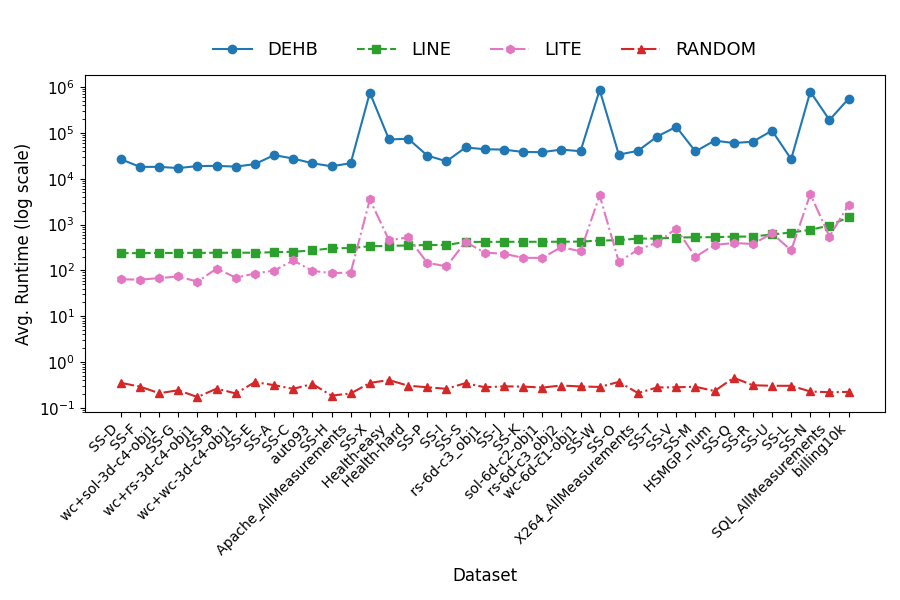}
        \caption{Runtime comparison (log scale, milliseconds) of different optimizers across datasets.}
        \label{fig:perf}
 \vspace{-3mm}       
\end{figure*}
Figure~\ref{fig:perf} shows the runtimes of our different methods.
Our simplest pool-based method (RANDOM)
that just samples the buckets, runs 10,000 times faster than the membership query method.
LITE and LINE are slower, but still run 100 times faster than DEHB. Hence we say:
\begin{conclusion}{ RQ3) Is  {\sffamily BINGO}-based reasoning fast?}
 Yes. RANDOM is the fastest but, given the Figure~\ref{pcf} results, when dealing with less than 50 labels, we recommend LINE or LITE. 
\end{conclusion}
The vertical axis of Figure~\ref{fig:perf} is in milliseconds so it could be argued,
in terms of absolute values, these 
are all similar. We would dispute that, saying that our evaluation times here  are near instant (just look at the labels). In practice, each evaluation could be very slow. For example, to check a proposed configuration for a Makefile, we may need to recompile an entire system, then run an extensive test suite.
In that context, the time differences in Figure~\ref{fig:perf} would become of great practical importance. 

 The smaller a code base, the faster newcomers can understand, adopt, apply, and adapt that code. 
 Hence it is important to note that 
 the code required to implement all our methods is orders of magnitude different in size.  
RANDOM is literally one line. LITE is 25 lines long (assuming a Naive Bayes classifier is available elsewhere). And even if it is implemented from scratch, the K-Means++ centroid finder is just 20 lines (in Python). DEHB,  on the other hand, is a hundred times bigger\footnote{Measured from \url{https://github.com/automl/DEHB/blob/master/src/} via  \newline
{\tt cat  src/*/*/*.py  |wc} 
$\rightarrow$ 2404 lines }. Hence:
\begin{conclusion}{ RQ4) Is  {\sffamily BINGO}-based reasoning simple?}
 Yes. Often, simple RANDOM selection suffices (but when data labels are scarce, we recommend   LITE and LINE-- which are not much larger)
\end{conclusion}

 \section{Discussion}
\subsection{Threats  to Validity}\label{tov}

As with any empirical study, biases can affect the final
results. Therefore, any conclusions made from this work must
be considered with the following issues in mind:
 
{\em Evaluation bias:} This paper uses one measure of success called {\em distance to heaven}. While we feel justified in our  
use of this metric  (see  \S\ref{eval})
it is true that other evaluation biases could change all our  results.  To check
that, we have repeated the above analysis using another evaluation bias:
the Chebyshev distance to utopia measured preferred by MOEA/D researchers~\cite{zhang2007moea}. While this changed some of the numbers seen above,
all the answers to our research questions remained the same.

{\em Order bias:} As with any stochastic analysis, 
the results can be effected by the ordering in which
the examples are randomly selected.   To mitigate this order bias, we run
the experiment 20 times, each time using a different seed for the random number generator.

{\em Learner bias:} This paper has compared pool-based inference to member-ship query methods. The case for that comparison was made above but, as with any empirical study, some new algorithm could arrive tomorrow that changes all these results. For example, here we use DEHB and 
there exists one potentially better successor to DEHB, called  PriorBand~\cite{Mallik23}.
While the underlying principles of  PriorBand are  applicable to many optimization problems,  PriorBand is specifically tailored for deep learning applications (which are not studied here). Hence, for this study, we chose to remain with DEHB. 

In our view, {\em sampling bias} is the biggest threat to the validity of this paper. 
Our reading of this literature is that most papers evaluate themselves with five data sets or  less (sometimes, only one) so the sample size used here ($N=39$) is far larger than most other papers. That said, this paper does not explore
data from the test case generation or test case reduction literature. This
is clear a direction for future work. 

More generally, another complaint about our sample
might be that  our datasets are {\bf trivially small} \cite{yang2022survey},
especially  when compared to large SE textual/image/code data sets. In reply, we say our
Figure~\ref{combinedtable} datasets were used in top SE journals
like IEEE Trans. SE~\cite{peng2023veer, Chen19, nair2018finding}, Empirical
Softw. Eng.~\cite{peng2023veer, xia2020sequential,guo2018data}, and
ACM Trans. SE Methodologies~\cite{lustosa2024learning}, thus demonstrating
that  our research community finds this kind of data interesting.
Also,
our dataset sizes match HPO benchmarks (e.g.,
NAS-HPO-Bench: $\le$27,438 configs; HPOBench: $>$90\% $<6$5k configs),
which is relevant as we use DEHB and our sizes are typical in HPO research.

Another complaint is that our datasets are
{\bf simplistic old-fashioned tabular} data. We argue tabular data is not
simple or outdated. Its processing is challenging, even for LLMs.
Somvanshi et al.~\cite{somvanshi2024survey} report that ``despite
deep learning’s success in image and text domains, tree-based
models like XGBoost and Random Forests continue to outperform neural
networks on medium-sized tabular datasets. This performance gap
persists even after extensive hyperparameter
tuning''~\cite{somvanshi2024survey}.

Tabular data remains widely used. Somvanshi
et al.~\cite{somvanshi2024survey} call it ``the most commonly used
data format in many industries...finance, and transportation.''.
Studies~\cite{ling2024trading, menzies2023best} show commercial use
in data synthesis/privatization for health/government. GitHub-scale
code analytics uses tabular data (e.g.,
CommitGuru\footnote{\url{http://commit.guru/}}, used in many SE
papers\footnote{See \url{http://tiny.cc/guruSince2019} for
CommitGuru usage.}). Software HPO uses data like
Table~\ref{combinedtable}; many tasks involve cloud software
HPO~\cite{peng2023veer}. Xia et al.~\cite{xia2020sequential}
analyzed 1600 GitHub projects using tables (13 indicators x 60
months). Non-SE HPO techniques
(Hyperband~\cite{li2018hyperband},
SMAC3~\cite{lindauer2022smac3}, DEHB~\cite{awadijcai2021p296})
also use tabular benchmarks.

\subsection{Future Work}

Certain     empirical quirks are central
to much of the current thinking in SE. 
 For example, operating systems algorithms are critically reliant on
 a {\em principle of locality}; i.e.
 computer programs tend to access the same set of data or instructions repeatedly over a short period (temporal locality) or access data items that are close to each other in memory (spatial locality)~\cite{silberschatz2018operating}.
Similarly, LLMs rely on their own empirical quirks:
 \bi
 \item Zipf's 
 {\em Law of Language} that the next token in a  stream can be predicted with high probability, given knowledge of $n$ prior tokens \cite{mellor2006psycho}.
 \item
 The success of the {\em attention mechanisms}~\cite{vaswani2017attention} where
  dynamic weighing of conditional probabilities between past tokens significantly improves predictions. \ei

This paper has reported another empirical quirk-- that data falls to just a few buckets. We conjecture that this effect might prove to be  just as important as locality,  Zipf's Law, or attention mechanisms.
Suppose some of  the focus and funding devoted
to LLMs was refocused on this
new  data quirk?
If so, would that mean:
\begin{itemize}
\item Data collection  would be simpler, needing less examples?
\item Test suites  would  shrink,
covering  a few
  in-out buckets?
\item Regression testing would be  faster/cheaper since we run fewer (and faster) test suites?
\item Stakeholders would explore more requirements, more quickly  since there the net effect of many  requirements  collapses to just a few buckets?
\item AI would use less  energy.  LLMs would not be used  for simple tasks (and many tasks will be simple)?
\item Data analytics would be faster and easier to reproduce?
\item Research results would be easier to replicate and refine?
\item Industrial tools  offer real-time feedback to analysts?
\item Auditing would only require inspecting a few bucketed samples;
e.g. small regression trees trained on ~50 samples would explain programs, very  well, since such trees are fast to read and audit?
\item  AI would be trusted, easier, cheaper, and widespread?
\item Stakeholders   find AI easier to understand and control?
\item New science   emerges from AI-fueled insight loops?
\item Human-AI teams   outperform either working alone?
\item In time, AI   enhance what it means to be human?
\item If the double BINGO effect holds, then all the above  can  happen in an instant?
\end{itemize} 
All these questions are worth studying,   in future research.

\section{Conclusion}

This study presents strong empirical evidence for a surprising and impactful phenomenon in software engineering optimization: the BINGO effect, wherein high-dimensional configuration spaces often collapse into a small number of densely populated buckets. By leveraging this structure, we demonstrate that lightweight, stochastic optimizers can achieve performance on par with state-of-the-art methods while requiring orders of magnitude less computational effort and far fewer labeled examples.

Our analysis spans 39 diverse SE optimization problems, consistently showing that simpler algorithms suffice when the data exhibits BINGO-like structure. We further propose practical heuristics for identifying when such simplifications are applicable. These findings challenge prevailing assumptions about the need for complex, resource-intensive optimization techniques in SE contexts.

Looking forward, this work encourages a shift in focus—from scaling up algorithmic complexity to understanding the underlying structure of real-world SE data. In many cases, optimization may not require more data, deeper models, or extensive computation—but rather, better recognition of inherent simplicity.

\IEEEtriggeratref{67}
\bibliographystyle{IEEEtran}
\bibliography{_sorted}

\end{document}